\documentclass[sigconf,nonacm]{acmart}

\usepackage{subfigure}

\begin{document}

\title{GRecX: An Efficient and Unified Benchmark for GNN-based Recommendation}

\author{Desheng Cai$^{1}$, Jun Hu$^{2}$, Quan Zhao$^{1}$}
\author{ Shengsheng Qian$^{2,3}$, Quan Fang$^{2,3}$, Changsheng Xu$^{2,3}$}
\affiliation{
  \institution{$^{1}$Hefei University of Technology}
  \country{}
  % \country{China}
}
\affiliation{
  \institution{$^{2}$National Laboratory of Pattern Recognition, Institute of Automation, Chinese Academy of Sciences}
  \country{}
  % \country{China}
}
\affiliation{
  \institution{$^{3}$School of Artificial Intelligence, University of Chinese Academy of Sciences}
  \country{}
  % \country{China}
}
% \affiliation{
%   \institution{$^{3}$Peng Cheng Laboratory, ShenZhen, China}
%   \country{}
%   % \country{China}
% }
\email{{caidsml, hujunxianligong}@gmail.com, zhaoquan219@163.com}
\email{ {shengsheng.qian, qfang}@nlpr.ia.ac.cn, csxu@nlpr.ia.ac.cn}

% \affiliation{%

% \affiliation{%

% \affiliation{%

\renewcommand{\shortauthors}{Desheng Cai and Jun Hu, et al.}

\begin{abstract}
In this paper, we present GRecX, an open-source TensorFlow framework for benchmarking GNN-based recommendation models in an efficient and unified way.
% , which is built with TensorFlow and tf-geometric. 
% building GNN-based recommendation models with TensorFlow and tf-geometric. 
%
GRecX consists of core libraries for building GNN-based recommendation benchmarks, as well as the implementations of popular GNN-based recommendation models.
The core libraries provide essential components for building efficient and unified benchmarks, including FastMetrics (efficient metrics computation libraries), VectorSearch (efficient similarity search libraries for dense vectors), BatchEval (efficient mini-batch evaluation libraries), and DataManager (unified dataset management libraries).
Especially, to provide a unified benchmark for the fair comparison of different complex GNN-based recommendation models, we design a new metric GRMF-X and integrate it into the FastMetrics component.
Based on a TensorFlow GNN library tf\_geometric~\footnote{https://github.com/CrawlScript/tf\_geometric}, GRecX carefully implements a variety of popular GNN-based recommendation models.
We carefully implement these baseline models to reproduce the performance reported in the literature, and our implementations are usually more efficient and friendly.
In conclusion, GRecX enables users to train and benchmark GNN-based recommendation baselines in an efficient and unified way.
We conduct experiments with GRecX, and the experimental results show that GRecX allows us to train and benchmark GNN-based recommendation baselines in an efficient and unified way.
The source code of GRecX is available at https://github.com/maenzhier/GRecX.

\end{abstract}

\maketitle

\begin{table}
% \small
	\centering
				% \vspace{-1mm}
		\caption{An Example of Experimental Settings of Different Baselines. NSS denotes negative sampling strategies, where "single" and "multiple" represent the number of negative samples. The comparison is unfair due to different settings.}
			\vspace{-1mm}
\scalebox{0.9}{
	\begin{tabular}{c|c|c|c} 
		\toprule 
		Models                       & NGCF                & LigthGCN         & UltraGCN$_{base}$\\
		\midrule                                                          
		Dim                          & 64                  & 64               & 64 \\
		Interactions                 & $\checkmark$        & $\checkmark$     & $\checkmark$ \\
		Social                       & $\times$            & $\times$         & $\times$ \\
		Pretrained                   & $\times$            & $\checkmark$     & $\times$ \\
		GNNs                         & $\checkmark$        & $\checkmark$     & $\times$          \\    
		Negtive Sampling Strategy    & single(1)           & single(1)        & \textbf{multiple(500+)}   \\
		Loss                         & BPR                 & BPR              & \textbf{BCE}            \\
% 		Num negative samples         & 1                 
		\bottomrule                                  
	\end{tabular}
	}
	\label{tab:baselines-features}
	\vspace{-2mm}
\end{table}

% \begin{table}
% % \small
% 	\centering
% 		\caption{An Example of Experimental Settings of Different Baselines. NSS denotes negative sampling strategies, where "single" and "multiple" represent the number of negative samples. The comparison is unfair due to different settings.}
% 			\vspace{-3mm}
% \scalebox{0.80}{
% 	\begin{tabular}{c|c|c|c|c|c|c} 
% 		\toprule
% 		Models          & MF                & Diffnet            & Diffnet++          & NGCF              & LigthGCN       & UltraGCN\\
% 		\midrule
% 		Dim              & 64               &  64                &  64                & 64                & 64             & 64 \\
% 		Interactions     & \checkmark       & \checkmark         & \checkmark         & \checkmark        & \checkmark     & \checkmark \\
% 		Social           & \times           & \checkmark         & \checkmark         & \times            & \times         & \times \\
% 		Pretrained       & \times           & \checkmark         & \checkmark         & \times            & \checkmark     & \times \\
% 		GNNs             & \times           & \checkmark         & \checkmark         & \checkmark        & \checkmark     & \times          \\    
% 		NSS              & single           & single             & single             & single            & single         & multiple   \\
% 		Loss             & BCE              & BPR                & BPR                & BPR               & BPR           & BCE            \\
% 		\bottomrule
% 	\end{tabular}
% 	}
% 	\label{tab:baselines-features}
% 	\vspace{-3mm}
% \end{table}

\section{Introduction}
Personalized recommendation is an important yet challenging task, which has attracted substantial attention in the past decade.
Most traditional approaches consider recommendation as a matching task~\cite{sigirXuHL18}, and can be solved by estimating the matching score based upon semantic representations of users and items~\cite{wwwHeLZNHC17, DBLP:conf/mm/HuangFQSLX19, DBLP:conf/aaai/JiZWSWTLH21}.
Recently, graph representation learning approaches are emerging tools to pursue a meaningful vector representation for each node in graphs, which can effectively model users, items and their corresponding relationships.
Graph Neural Networks (GNNs), such as GCN \cite{DBLP:conf/iclr/KipfW17}, GraphSAGE \cite{DBLP:conf/nips/HamiltonYL17} and GAT \cite{DBLP:conf/iclr/VelickovicCCRLB18}, have shown impressive performance in aggregating feature information of neighboring nodes.
In recommender systems, the interactions between users and items can be represented as a bipartite graph and the goal is to predict new potential edges (i.e., which items could a user be interested in), which can be achieved with GNNs, which are called GNN-based recommendation methods.
GNN-based recommendation techniques has attracted researchers and engineers from a variety of fields, and they have been utilized to build various real-world applications such as medicine recommendation~\cite{DBLP:conf/icde/JinZ00W20}, micro-video recommendation~\cite[]{DBLP:conf/mm/WeiWN0HC19, Cai2021HeterogeneousHF}, and social recommendation~\cite[]{DBLP:conf/bigdataconf/BaiZWN20, DBLP:conf/pricai/ZhangLW18, DBLP:conf/sigir/YangWHZW21}.

Although many existing approaches provide official implementations, it is still difficult to build efficient and unified benchmarks due to the following limitations:
\textbf{(1)} The evaluation of GNN-based recommendation approaches may involve many computationally expensive operations, which should be optimized to perform efficient benchmarking.
However, most official implementations usually ignore the problem, and some of the computationally expensive operations are still widely used by these implementations.
For example, these implementations usually rely on the matrix multiplication between the user representation matrix and the item representation matrix to perform the topK retrieval of items for users, which may result in large time and space complexity.
\textbf{(2)} It is difficult to build a unified benchmark since different baselines may adopt different loss functions, different negative sampling strategies, etc. 
For example, UltraGCN~\cite{DBLP:conf/cikm/MaoZXLWH21} adopts NGCF~\cite{DBLP:conf/sigir/Wang0WFC19} and LightGCN~\cite{DBLP:conf/sigir/0001DWLZ020} as baselines.
However, the negative sampling strategy of UltraGCN~\cite{DBLP:conf/cikm/MaoZXLWH21} is different from that of NGCF~\cite{DBLP:conf/sigir/Wang0WFC19} and LightGCN~\cite{DBLP:conf/sigir/0001DWLZ020}.
As shown in Table~\ref{tab:baselines-features}, NGCF and LightGCN adopt a negative sampling strategy that uses only one negative sample, while negative sampling strategy of UltraGCN$_{base}$, which represents the standard version of UltraGCN, samples more than 500 negative samples. 
It is well known that GNN-based recommendation models can be improved by simply increasing the number of negative samples~\cite{DBLP:conf/icde/JinZ00W20}.
As a result, the experimental results reported by UltraGCN~\cite{DBLP:conf/cikm/MaoZXLWH21} cannot verify the effectiveness of the model.
Note that we implement the UltraGCN$_{base}$ model as a baseline.

% as shown in Table~\ref{tab:baselines-features},  

% Table~\ref{tab:baselines-features} shows the differences between several baselines.
% %
% For example, although Diffnet/Diffnet++ and LightGCN are popular GNN-based recommendation models, they are not comparable due to different datasets and GNN layers.
% %
% LightGCN and UltraGCN are both based on user-item interaction graphs for recommendations but use different GNN layers, loss functions, and negative sampling strategies.
% %
% Therefore, for GNN-based recommendation models, we cannot determine whether GNN layers of LightGCN are better than Diffnet/Diffnet++'s, or the components of UltraGCN are superior over LightGCN's.
%
% Note that in table~\ref{tab:baselines-features}, NSS is represented as negative sampling strategies. When losses are calculated, "single" means one negative sample for per positive sample, and "multiple" means multiple negative samples for per positive sample.

In this paper, we present GRecX, an open-source TensorFlow framework for benchmarking GNN-based recommendation models in an efficient and unified way.
To address the efficiency problem, we develop core libraries to provide essential components for build efficient benchmarks, including FastMetrics (efficient metrics computation libraries), VectorSearch (efficient similarity search libraries for dense vectors), BatchEval (efficient mini-batch evaluation libraries), and DataManager (unified dataset management libraries).
To build a unified benchmark for the fair comparison of different complex GNN-based recommendation models, we design a new metric named GRMF-X and integrate it into the FastMetrics component.
In addition, we also provide efficient implementations of a variety of popular GNN-based recommendation models, which enable us to build a more comprehensive benchmark.
We conduct experiments with GRecX, and the experimental results show that GRecX allows us to train and benchmark GNN-based recommendation baselines in an efficient and unified way.
%
% The source code of GRecX is available at https://github.com/maenzhier/GRecX.
%
All features of GRecX and a collection of examples are  provided with the source code, which is available at https://github.com/maenzhier/GRecX.

% Despite the performance improvement of GNN-based recommendation models,  we 

% In this paper we present GRecX, an open-source Python library for building GNN-based recommendation models with TensorFlow and the tf-geometric software. 
% %
% GRecX implements a large set of methods for recommendation models on graphs, including user-item recommendations and social recommendations, as well as utilities for processing graphs and loading popular benchmark datasets.

% %
% All features of GRecX and a collection of examples is provided with the source code.
% The project is released on GitHub under MIT license.

\begin{table*}[!tbpb]
	\centering
	\vspace{-2mm}
		\caption{Statistics of Datasets from LightGCN.}
			\vspace{-2mm}
	\begin{tabular}{c|c|c|c|c} 
		\toprule
		Dataset        & User          & Item        & Interaction    & Density \\
		\midrule
		yelp2018       & 31,668        & 38,048      & 1,561,406      & 0.00130  \\
		gowalla        & 29,858        & 40,981      & 1,027,370      & 0.00084   \\
		amazon-book    & 52,643        & 91,599      & 2,984,108      & 0.00062  \\
		\bottomrule
	\end{tabular}
	\label{tab:datasets}
	\vspace{-2mm}
\end{table*}

\begin{table*}[tp!]
% \small
	\centering
	\setlength{\tabcolsep}{4 pt}
		\caption{Performance of Baselines on Two Datasets in terms of the BPR Loss.}
		\vspace{-2mm}
% 		\scalebox{0.9}{
	\begin{tabular}{c|c||c|c|c|c|c|c|c} 
		\toprule
		\multicolumn{2}{c|}{\textbf{Parameter Settings}} & \textbf{MF} &\textbf{UltraGCN$_{base\_s}$}  &\textbf{LightGCN} & \textbf{UltraGCN$_{base}$} & \textbf{MLP+MF}   & \textbf{NGCF}    & \textbf{MF}      \\
			    \midrule
	    \multicolumn{2}{p{24mm}||}{Number of Negative Samples} 
	                                 & \multicolumn{3}{c|}{1}     
                                     & \multicolumn{1}{c|}{800}                                            
	                                 & \multicolumn{3}{c}{1} \\
	    \midrule
	    \multicolumn{2}{p{24mm}||}{Dimensionality} 
	                                 & \multicolumn{4}{c|}{64}                                            
	                                 & \multicolumn{2}{c|}{$256 (64 \rightarrow 256)$}
	                                 & \multicolumn{1}{c}{$256$} \\
	    \midrule
	    
		Datasets     & Metrics                  & \multicolumn{7}{c}{}      \\
	    \midrule
	    
		yelp2018     & NDCG@20                  & 0.04938     &    ---                & 0.0524           &    ---       & 0.0493    & 0.0484  & 0.0522 \\
		             & GRMF-X(\%)               & 0.0\%       &    ---                &  6.12\%          &    ---       & -0.77\%  &-1.98\%  & 5.71\%       \\
	   \midrule
		gowalla      & NDCG@20                  & 0.1400      &    ---                & 0.1485           &    ---       & 0.1388   & 0.1394  & 0.1477 \\
		             & GRMF-X(\%)               & 0.0\%       &    ---                & 6.07\%           &    ---       & -0.86\%  & -0.42\% & 5.36\% \\
	   \midrule
		amazon-book  & NDCG@20                  & 0.0265      &    ---                & 0.0309           &    ---       & 0.0260   & 0.0268   & 0.0310 \\
		             & GRMF-X(\%)               & 0.0\%       &    ---                & 16.60\%          &    ---       & -1.89\%  & 1.13\%   & 16.98\% \\
		\bottomrule
	\end{tabular}
% 	}
	\label{tab:baselines-res-bpr}
	\vspace{-2mm}
\end{table*}

\begin{table*}[tp!]
% \small
	\centering
	\setlength{\tabcolsep}{4 pt}
		\caption{Performance of Baselines on Two Datasets in terms of the BCE Loss.}
		\vspace{-2mm}
% 		\scalebox{0.9}{
	\begin{tabular}{c|c||c|c|c|c|c|c|c} 
		\toprule
		\multicolumn{2}{c|}{\textbf{Parameter Settings}} & \textbf{MF} &\textbf{UltraGCN$_{base\_s}$}  &\textbf{LightGCN} & \textbf{UltraGCN$_{base}$} & \textbf{MLP+MF}   & \textbf{NGCF}    & \textbf{MF}      \\
			    \midrule
	    \multicolumn{2}{p{24mm}||}{Number of Negative Samples} 
	                                 & \multicolumn{3}{c|}{1}     
                                     & \multicolumn{1}{c|}{800}                                            
	                                 & \multicolumn{3}{c}{1} \\
	    \midrule
	    \multicolumn{2}{p{24mm}||}{Dimensionality} 
	                                 & \multicolumn{4}{c|}{64}                                            
	                                 & \multicolumn{2}{c|}{$256 (64 \rightarrow 256)$}
	                                 & \multicolumn{1}{c}{$256$} \\
	    \midrule
	    
		Datasets     & Metrics                  & \multicolumn{7}{c}{}      \\
	    \midrule
	    
		yelp2018     & NDCG@20                  & 0.0471      &   0.04776             & 0.04587          &    0.05537       & 0.0456   & 0.03955        & 0.0515 \\
		             & GRMF-X(\%)               & 0.0\%       &   1.40\%              & -2.61\%          &    17.56\%       & -3.18\%  &-16.08\%        & 9.34\%       \\
	   \midrule
		gowalla      & NDCG@20                  & 0.1298      &    0.1387             & 0.1300           &    0.1482        & 0.1361   & 0.1228         & 0.1480\\
		             & GRMF-X(\%)               & 0.0\%       &    6.86\%             & 0.15\%           &    14.18\%       & 4.85\%   & -5.39\%         & 14.02\% \\
	   \midrule
		amazon-book  & NDCG@20                  & 0.0258      &    0.0319             & 0.0300           &    0.0350        & 0.0255   & 0.0264   & 0.0312 \\
		             & GRMF-X(\%)               & 0.0\%       &    23.64\%            & 16.28\%          &    35.66\%       & -1.18\%  & 2.33\%   & 20.93\% \\
		\bottomrule
	\end{tabular}
% 	}
	\label{tab:baselines-res-bce}
	\vspace{-2mm}
\end{table*}

\section {Overview}

\begin{figure}[tbp!]
	\centering
% 	\vspace{-3mm}
% 	\includegraphics[width=3.2in]{./framework.pdf}
	\includegraphics[width=3.2in]{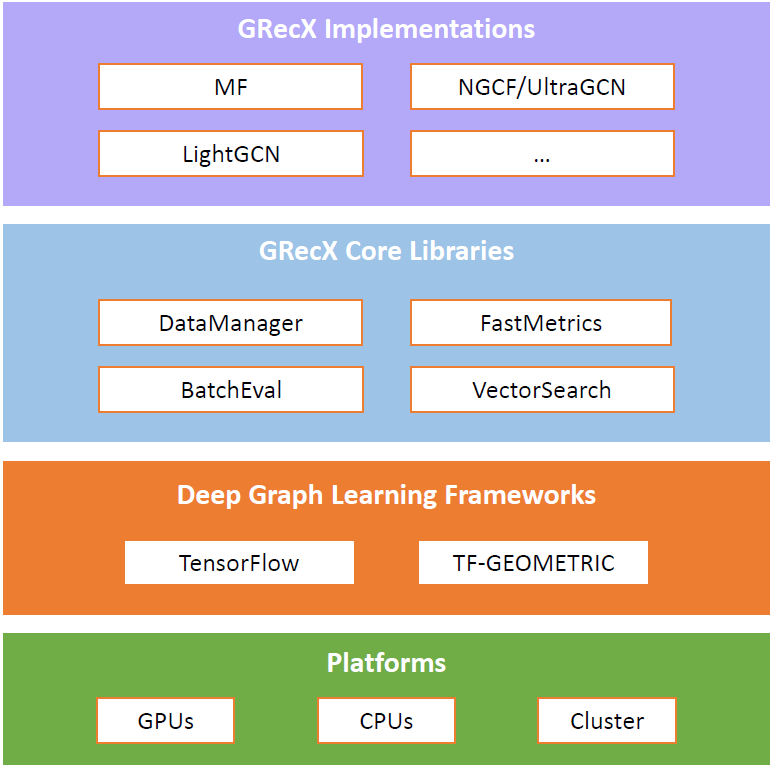}
	\vspace{-3mm}
	\caption{Overview Framework of GRecX}\label{fig:framework}
	\vspace{-3mm}
\end{figure}

Figure~\ref{fig:framework} shows the overall framework of GRecX, which mainly consists of core libraries and implementations of popuplar GNN-based recommendation models.
In this section, we provide an overview of the framework of GRecX.

\subsection{Core Libraries}

The core libraries provide essential components including FastMetrics (efficient metrics computation libraries), VectorSearch (efficient similarity search libraries for dense vectors), BatchEval (efficient mini-batch evaluation libraries), and DataManager (unified dataset management libraries).
There components enable us to build efficient and unified benchmarks. 
In this section, we will introduce each component of GRecX's core libraries in detail.

\subsubsection{FastMetrics}\label{sec:fast_metrics}
FastMetrics provides efficient implementations for various widely-used recommendation metrics, such as NDCG@N, Precision, and Recall.
It is non-trivial to implement efficient metrics computation libraries for recommendation due to the complexity of real-world data.
Moreover, for fair comparison, we design a new metric named GRMF-X and integrate it into FastMetrics.
GRMF-X is the abbreviation for \textbf{G}ain \textbf{R}elative to \textbf{MF} in terms of the metrics \textbf{X}, and it is defined as follows:
\begin{equation}\label{eq:GRMF-X}
    GRMF-X(MODEL, CTX) = \frac{X\_SCORE(MODEL, CTX)}{X\_SCORE(MF_{tuned}, CTX)} - 1.0
\end{equation}
where $X\_SCORE(MODEL, CTX)$ is the evaluated score of model $MODEL$ under the metric $X$ and context $CTX$, and $MF_{tuned}$ denotes a tuned Matrix Factorization (MF)~\cite{Koren2009MatrixFT} model.
We introduce this naive evaluation metric since it can effectively help us to verify the superiority of GNN-based recommendation models.

There are mainly two reasons why our unified benchmark can benefit from GRMF-X:
(1) We observe that although some research's experimental results show that their proposed model outperforms all the baselines, the reported performance of the baselines or the proposed models may not be competitive with the simple well-tuned MF model.
Note that although some research provide the performance of MF, they may employ a MF model that is not well tuned, which usually show poor performance.
This shows that the authors do not conduct experiments with well-implemented baselines, and thus the experimental results are not convincing.
(2) We also observe that some research do not conduct experiments of different baselines under the same context.
Here we use context $CTX$ to denote the some important settings beyond the GNN architectures, such as the negative sampling strategies.
For example, as shown in Table~\ref{tab:baselines-features}, NGCF and LightGCN employ a negative sampling strategy that uses only one negative sample, while UltraGCN's negative sampling strategy samples more than 500 negative samples.
It is well known that GNN-based recommendation models can be improved by simply increasing the number of negative samples~\cite{DBLP:conf/icde/JinZ00W20}; therefore, directly adopting the performance of the official implementations may result in unfair comparison.
In Equation~\ref{eq:GRMF-X}, we implicitly constrain that different models are compared under the same context $CTX$.
Note some hyper-parameters such as the learning rate and L2 coefficient are not considered as the context.
For these hyper-parameters, different models may rely on different parameter settings to achieve their best performance, and we should carefully tune these hyper-parameters to obtain the best performance.

\subsubsection{VectorSearch}
Many GNN-based recommendation approaches perform recommendation by similarity based searching, which ranks items based on the similarity scores between the dense vector representation of users and items.
Most evaluation metrics require a global scan over all the items, which may result in large time and space complexity.
We leverage industrial solutions such as Faiss~\cite{DBLP:journals/tbd/JohnsonDJ21} to build efficient similarity search libraries for dense vectors named VectorSearch, which can perform dense vector searching efficiently with millions of candidate items.
Furthermore, we improve the VectorSearch module with several re-implemented similarity search methods, such as binary vectors and compact quantization codes, to make searching more efficient and flexible.

\subsubsection{BatchEval}

The evaluation of GNN-based recommendation models can benefit from mini-batch techniques, which can take advantage of GPU's parallel processing ability to improve the efficiency of the evaluation.
It requires a lot of tricks to design mini-batch implementations on irregular real-world recommendation data.
To provide friendly and handy mini-batch solutions, we design BatchEval.
The users only need to provide the learned representations or the similarity computation function, and BatchEval can automatically perform efficient mini-batch based evaluation with the provided information.

\vspace{-2mm}

\subsubsection{DataManager}
DataManager provide abstract Dataset class as interfaces for users to custom handy dataset APIs.
The abstract Dataset class can automatically handle the whole lifecycle of data processing, such as data downloading, data preprocessing, data caching, etc.
Usually, users can easily custom their Dataset class by subclassing the abstract Dataset class, providing the download urls of raw dataset, and overriding the preprocessing process.
Then, the abstract Dataset class will handle the rest of data processing process.
In addition, we already implement several widely-used recommendation datasets as Dataset classes, which can be directly used to load these datasets.

\vspace{-3mm}

\subsection{Implementations of GNN-based Recommendation models}
In this paper, we implement a basic MF model~\cite{Koren2009MatrixFT} and the state-of-the-art GNN-based recommendation algorithms (e.g. NGCF~\cite{DBLP:conf/sigir/Wang0WFC19}, LightGCN~\cite{DBLP:conf/sigir/0001DWLZ020}, UltraGCN$_{base}$~\cite{DBLP:conf/cikm/MaoZXLWH21}) as baselines.
Especially, we carefully implement these baseline models to reproduce the performance reported in the literature, and our implementations are usually more efficient and friendly.

\section{Experiments}

We conduct experiments on several benchmark datasets to provide a fair comparison of some popular baselines with GRecX.
The complete experimental results are at https://github.com/maenzhier/GRecX.
We are still working on some baselines to provide more comprehensive benchmark.
Here, we only list partial experimental results on two benchmark datasets to demonstrate the performance of models implemented by GRecX and our new evaluation metrics GRMF-X.

% \vspace{-1mm}
\subsection{Datasets and Baselines}

We use three datasets: yelp2018, gowalla, and amazon-book.
% We use two datasets: yelp2018 and gowalla
%
Note that some dataset may have different versions, and we use the version used by LightGCN~\cite{DBLP:conf/sigir/0001DWLZ020}.
The statistics of the three datasets is listed in Table~\ref{tab:datasets}.
In terms of the baselines, we choose a basic model MF~\cite{Koren2009MatrixFT}, and three state-of-the-art GNN-based recommendation models UltraGCN$_{base}$~\cite{DBLP:conf/cikm/MaoZXLWH21}, NGCF and LightGCN~\cite{DBLP:conf/sigir/0001DWLZ020}.

\begin{figure}%
  \centering
  \subfigure{
  \includegraphics[width=0.45\textwidth]{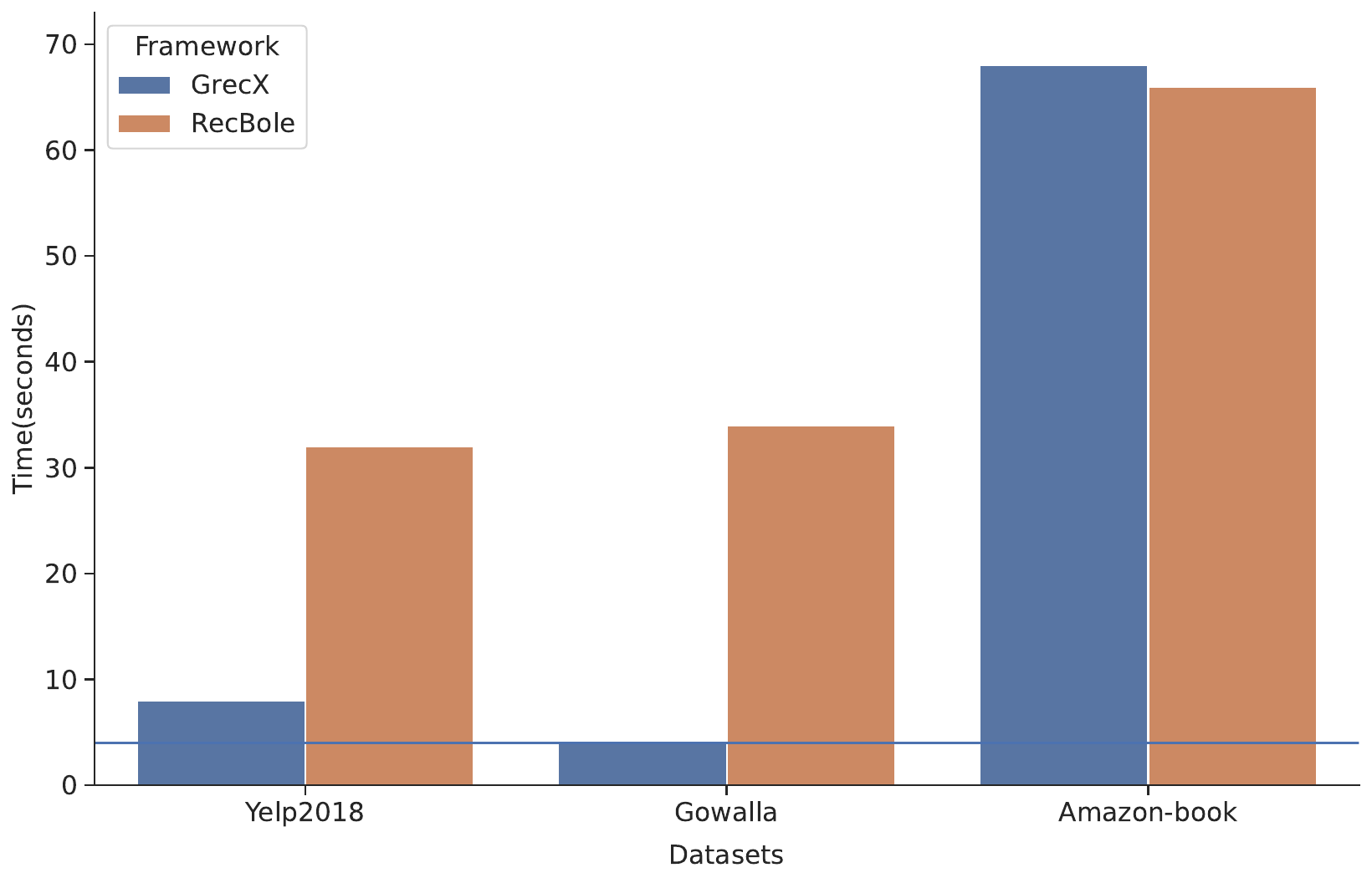}} 
  \subfigure{
  \includegraphics[width=0.45\textwidth]{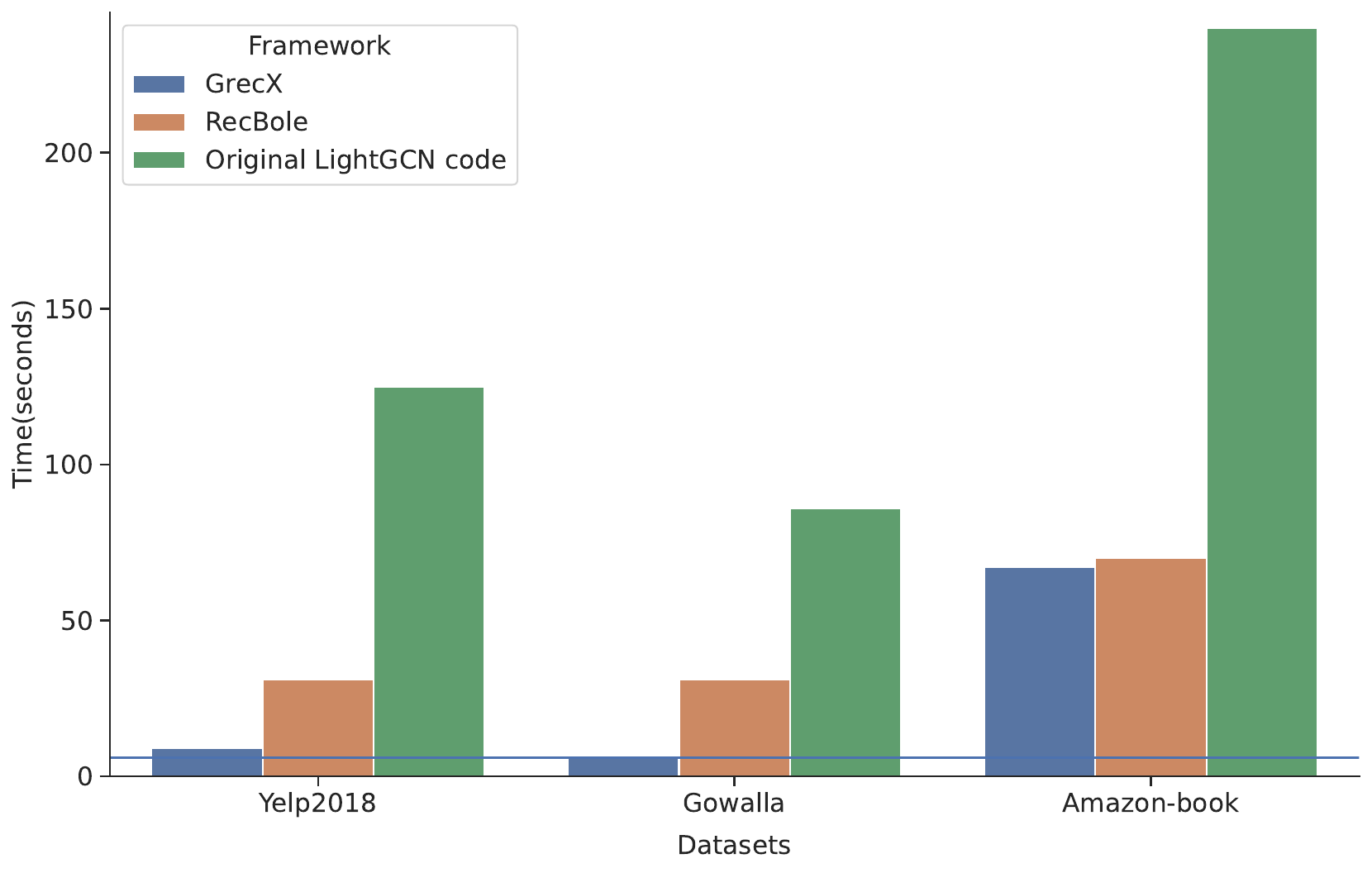}}
  \vspace{-1mm}
  \caption{LightGCN model inference time(seconds) comparison in term of different libraries on three datasets (\textbf{training\_rate=0.8}).}
  \vspace{-3mm}
 \label{fig:time-comparison}
\end{figure}

% \begin{figure}[t]%
%   \centering
%   \includegraphics[width=0.47\textwidth]{fig/lightGCN_grecx_recbole.pdf}
%   \vspace{-3mm}
%   \caption{LightGCN model inference time(seconds) comparison in term of different libraries on three datasets (\textbf{training\_rate=0.8}).}
%   \vspace{-3mm} 
%  \label{fig:time-comparison}
% \end{figure}

\subsection{Evaluation Metrics and Parameter Setting}

Here we employ a widely-used metrics NDCG@20 and our new metrics GRMF-X (GRMF-NDCG@20) for evaluation.
As mentioned in Section~\ref{sec:fast_metrics}, for other hyper-parameters, we carefully tune these hyper-parameters and report the performance with them.

We use BCE and BPR as the ranking losses respectively and experimental results are shown in Table~\ref{tab:baselines-res-bpr} and Table~\ref{tab:baselines-res-bce}.
For all comparable models, we set different parameters including number of negative samples and dimensionality of representation.
In terms of number of negative samples, we set it to 1 and 800 to align with the original experimental settings of NGCF (one negative sample), LightGCN (one negative sample) and UltraGCN$_{base}$ (800 negative samples).
Note that UltraGCN$_s$ means the UltraGCN$_{base\_s}$ model with using one negative sample in the training phase.
For the dimensionality of representation setting,  we set the dimensionality of learned user/item representations to 64 for all the models.
Specifically, we find that NGCF model concatenates the output embedding of each graph convolution layer including input layer to construct the final users/items' representations, which are then combined with the CF mechanism for recommendations.
Taking  NGCF model with three layers as an example, its dimensionality of final output representations is equal to 256 ($3 * 64 + 64$),  which may be unfair to other models.
So, we design MLP+MF model which just replaces all graph convolution layer of NGCF model with an MLP layer.
And we also show results of MF model with 64- and 256-dimensional representations.

\subsection{Performance}
In this paper, we implement a simple well-tuned MF model as a important baseline.
And we introduce a naive evaluation metric GRMF-X for verifying the superiority of GNN-based recommendation models effectively.
The results for recommendations on two datasets in terms of the BCE loss and BPR loss are reported in Table \ref{tab:baselines-res-bce} and \ref{tab:baselines-res-bpr}, respectively. 
Note that the experimental results are preliminary and will be updated continuously.

% From the results, the reported performance of our implementations may slightly be lower than the reported performance in the official implementations.
% %
% This is because we consider the efficient problem.
% %
% For example, for the LightGCN (BPR) on the yelp2018 dataset, either official implementations baselines or our implementation may spend more than 10 hours to improve NDCG@20 scores from 0.0524 to 0.0530 (the official result).
% %
% By ignoring such insignificant performance improvement (0.0006), the training time of our model can be dramatically reduced to only one or two hours.
%

%
In addition, compared with results of UltraGCN$_{base\_s}$ in table~\ref{tab:baselines-res-bce}, results of LightGCN with a same negative sampling startegy in table~\ref{tab:baselines-res-bpr} shows better performances.
Interestingly, compared with results of NGCF in table~\ref{tab:baselines-res-bpr} and table~\ref{tab:baselines-res-bce}, MF model (64- and 256-dimensional representations) and MLP+MF model achieve better results on three datasets in most cases.

Moreover, we conduct experiments on three datasets to analyze the time of model inference in terms of different LightGCN implement codes, including GrecX, RecBole, and original LightGCN codes. Experimental results are shown in Fig~\ref{fig:time-comparison}. 
From the histogram results, we can clearly see that, compared to RecBole code and original lightGCN code, the Inference efficiency of the lightGCN code implemented by our GrecX framework has been significantly improved, especially on the two datasets Yelp2018 and Gowalla.

\section{Conclusions}

In this paper, we present GRecX, an open-source TensorFlow framework for benchmarking GNN-based recommendation models in an efficient and unified way.
GRecX consists of core libraries for building GNN-based recommendation benchmarks, as well as the implementations of popuplar GNN-based recommendation models.
With GRecX, we can efficiently perform fair comparison between different GNN-based recommendation models in a unified benchmark.
In the future, we will integrate more baselines into GRecX and further improve the performance of both the core libraries and implementations.

\bibliographystyle{ACM-Reference-Format}
\bibliography{GRecX}

% \appendix









\end{document}